# Smart City Intelligent System
## Traffic Congestion Optimization using Internet Of Things

Kunal Verma and Vishal Paike
Department of Electronics and Electrical Engineering,
Indian Institute OF Technology, Guwahati-781039, Assam, India
v.kunal@iitg.ernet.in, paike@iitg.ernet.in

**Abstract**

The raising level of traffic imposes a great demand in the growth of intelligent traffic systems. With increase in complexity of alleviation, finding solutions to traffic congestion problem have become one of the challenges. Various optimization techniques have been proposed in literature to meet these challenges. This paper surveys different optimization techniques based on heuristics for automated traffic congestion control. Moreover, an approach based on River Formation Dynamics scheme is introduced to analyze the optimization problem for traffic congestion control and a scheme to extract real time information through Internet of Things is presented for superior efficiency and productivity.

## 1   Introduction

Road traffic conditions has taken the worse shape all around the globe in the recent times. As per the results of a survey done in India, the average number of vehicles are growing at a rate of 10.16% annually since last five years. This has resulted in a sudden increases in the number of vehicles on the roads, especially in the metropolitans. Spending hours in traffic jam has become a part and parcel for the metropolitan life style, leading to enormous economic, health and environmental hazards. The vehicle penetration in cities like Mumbai is suffering from about 590 vehicles per kilometres of road stretch and Bangalore with around 5 millions of vehicles ply over a network that extends barely up to 3000 kilometres. The concept of Intelligent Traffic System (ITS) has evolved as a solution to the outrageous problem of traffic congestion. There are sensors applied across the road units which tracks the vehicular behaviour with time and synchronises with the traffic flow control to minimise the congestion that may occur in future. Many researchers have proposed various methods to understand the dynamic behaviour of traffic flow using different algorithms. While most of these approaches uses the concept of evolutionary algorithms accounting for the randomness and mass level learning behaviour of traffic population, many researchers have also used the concepts of heuristics to analyse the traffic flow and determine the congestion density of a given road unit. This predicted value act as an indicator of any future traffic congestion that may occur and optimization techniques are taken into consideration to control these future congestion. This paper reviews various evolutionary and heuristic algorithms to analyse the traffic flow and to forecast the congestion. The concept of River Formation Dynamics to analyse the traffic congestion has been introduced along with an introduction to the deployment techniques using IoT.

## 2   Evolutionary Algorithms and Heuristic Models

### 2.1   Evolutionary Algorithms

Evolutionary Algorithms (EAs) are inspired by the biological model of evolution and natural selection first proposed by Charles Darwin in 1859. In natural world, evolution helps species adapt to their environment depending upon several external factors such as climate, availability of food and danger of predators. The solution to traffic congestion has also been sought as a self-evolution process by many researchers and several branches of EAs has been explored for the same. Various EAs such as the Genetic Algorithms (GAs), Particle Swarm Optimization (PSO), Artificial Neural Networks (ANN) and several Heuristic models has been explored by many researchers in this regard. The GAs in traffic congestion optimization is a metaheuristic approach similar to Darwin's evolution and optimization technique which uses the concepts of crossover and mutations of chromosomes. This method considers performance metrics such as queue length and vehicle idling time and stores it in a 1-dimensional array, the maximum value of which is the objective function to be minimised. The average values of these response variables enhances the evolution of the performance of the signalization plan in action. The PSO branch of EAs is a population based stochastic optimization technique developed by Dr. Eberhart and Dr. Kennedy in 1995, inspired by the social behaviour of bird flocking or fish schooling. Compared to GAs, PSO is easy to implement and has a few parameters to adjust. PSO simulates the behaviour of bird flocking. For a group of birds randomly searching for a single piece of food without knowing its exact positioning but estimate how far the food is after each estimation. As the best strategy, the effective one follows the bird which is nearest to the food. Unlike GAs, where the whole population moves like a group, PSO is a one way information sharing mechanism where evolution looks for the best solution. A further classification of SPO has been exploited for the traffic congestion analogy by many researchers. Artificial Bee Colony Optimization algorithms (ABC algorithm) is an optimization algorithm based on the intelligent behaviour of honey bee. Swarm ABC algorithm is used for optimising multivariable functions using the concepts of Ride Sharing and Ride Matching in traffic flow. Ride sharing technique assumes the participation of two or more person together sharing a vehicle when travelling from a few origins to few destinations, which reduces the number of trips. All the artificial bees are located in the hive at the beginning of search process, during which they communicates directly. Each artificial bee makes a series of local moves and in this way incrementally constructs solution to the problem. One another important class of SPO is the Ant Colony Optimization (ACO). ACO takes inspiration from the foraging behaviour of some ant species. These ants deposits pheromones on the ground in order to mark some favourable path that should be followed by other members of the colony. Using the Pheromone Communication Model (PCM), the traffic congestion density of a given section of road can be calculated. Another branch of the EAs that are most widely practiced are the Neural Networks. Back Propagation Neural Networks (BPNN) is the most mature and most widely used model for Traffic Flow Forecasting (TFF). BPNN is a multilayer feedback network and it follows supervised learning method. Programme design of BPNN involves designing network structure and network training. As an improvement over the error margin in BPNN, Simulated Annealing Genetic Back Propagation Algorithm for Traffic Flow Forecasting can get rid of local optimum and reduce the selection pressure. This model can increase the robustness of Genetic Back Propagation Algorithm and can reduce the errors of traffic flow prediction and increase the efficiency of TFF. Simulated Annealing (SA) is a method for solving unconstrained and bound constrained optimization problems. This method models the physical process of heating a material and then slowly lowering the temperature to decrease the defects, thus minimizing the system energy. At each iteration of SA algorithm, a new point is generated.

**Table 1:** Literature Survey on different approaches for traffic congestion optimization techniques.

| Title | Author/s | Milestones |
|---|---|---|
| Traffic Congestion Forecasting Algorithm based on Pheromone Communication Model | Satoshi Kurihara (Osaka University, Japan) | An improvement in the existing VICS and PCS model in China's Traffic System. |
| Traffic Congestion Evaluation and Signal Timing Optimization Based on Wireless Sensor Networks: Issues, Approaches and Simulations | WEIZHANG, GUO-ZHENTAN, NANDING AND GUANG-YUANWANG (School of Computer Science and Technology Dalian University of Technology) | Uses time-space model analysis of the Traffic System as an improvement over the existing state-space model. |
| Traffic Optimization System: Using a Heuristic Algorithm to Optimize Traffic Flow and Reduce Net Energy Consumption | Keshav Saharia (Monta Vista High School) | An optimization approach which considers the lighter as well as the heavier vehicles and assigns a congestion index to the controlling traffic light pairs. |
| Development of Stochastic Genetic Algorithm for Traffic Signal Timing Optimizations | Ahmed A. Ezzat, Hala A. Farouk, Khaled S. ElKilany, Ahmed F. Abdelmoneim (Department of Industrial & Management Engineering Arab Academy for Science, Technology & Maritime Transport | Has been tested on a section in Alexandria, Egypt. Uses non-linear objective function that takes majorly two measures, queuing length and vehicular waiting timings. |
| Solving Road-Network Congestion problems by a Multi Objective Optimization Algorithm using Brownian Agent Model. | Bin Jiang, Xiao Xu, Chao Yang, Renfa Li and Takao Terano. (College of Information Science and Engineerng, Hunan University, China.) | Agents are defined in complex or minimalistic ways. The agents works on a learning and optimization based algorithm. |
| Mitigating Traffic Congestion: Solving the Ride-Matching Problem by Bee Colony Optimization | Transportation Planning and Technology, April 2008 Vol. 31, No. 2, pp. 135 152 | An improvement over the Ride Sharing/Ride matching proposals and BCO to be a metaheuristic approach. |
| Traffic Signal Optimization in "La Almozara" District in Saragossa Under Congestion Conditions, Using Genetic Algorithms, Traffic Microsimulation, and Cluster Computing. (IEEE abstract) | IEEE TRANSACTIONS ON INTELLIGENT TRANSPORTATION SYSTEMS, VOL. 11, NO. 1, MARCH 2010 | Uses the three techniques viz. 1) genetic algorithms (GAs) for the optimization task; 2) cellular-automata based micro simulators for evaluating every possible solution for traffic- light programming times; and 3) a Beowulf Cluster, which is a multiple-instruction–multiple-data (MIMD) multicomputer of excellent price/performance ratio |

## 2.2 Random Walk based Heuristic Model

Another very useful Heuristic Algorithm that is widely used for the optimization problems is Random Walk. A Random Walk on the state-space is a sequence of states such that each state is chosen uniformly at a random from among the next states of the space. It uses much less space since it doesn't need to maintain any table to detect previously visited states. Also, it is inherently suitable for being parallelized. A Random Walk started in parallel on the same state-space from the start will explore more states than one Random Walk. This is because the probability of all n Random Walks taking the samepath decreases exponentially as the length of the path increases.

### 2.2.1 Markov Chain Heuristic Model

A Markov Chain is a process that consists of a finite number of states and some known probabilities of moving from one state to other. It is a random process that possess a property that is usually characterized as 'memorylessness'. The probability distribution of next step depends only on the current state and not on the sequence of the events preceded it. Further, Large Step Markov Chains are used as an improvement of Simulated Annealing. SA does not takes advantage of local opt heuristics. This means that instead of sampling local opt tours as does **L-K** repeated from random starts, the chain samples all tours. It enables restricting the sampling of Markov Chain to the local opt tours only. The bias then provided by the Markov Chain enables one to sample the shortest local opt tour more efficiently than local opt repeated starting randomly.

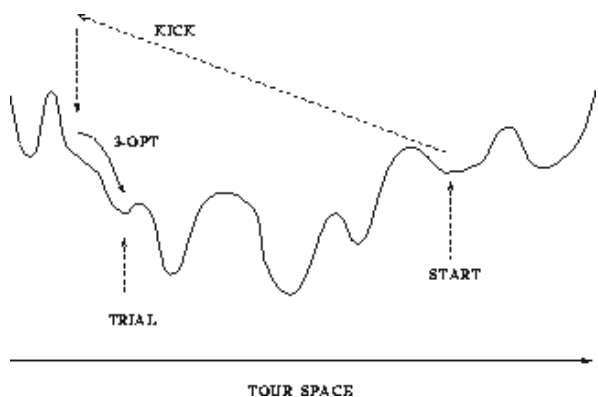

**Figure 1**: Schematic Representation of the Objective Function and of the Tour Modification Procedure Used in the Large-step Markov Chain.

## 3 River Formation Dynamics (RFD)

The pheromone trail in ACO tends to be stronger in short paths than in long paths owing to the time taken to be more in longer paths and subsequently the increases in the amount of pheromones at each point of the shorter path as it is traversed more. ACO method constructs good global paths by applying a local scope mechanism at each location. However, the ACO also suffers from some intrinsic problems. The pheromone paths may interfere and damage each other. Also, the local orientation may miss some simple graph peculiarities which would have allowed finding better paths. These two problems are a consequence of the following property of the method: When ants decide their next movement, only the pheromone trail at each possible destination is taken into account, but the pheromone trail at the origin is essentially ignored. If instead of this, the difference of trails between both places were considered, and this difference were required to be positive in order to choose a given edge as next step, then it would help to avoid both problems considered before. On the one hand, it is impossible that all edges in a cycle produce a positive increment, so cycles would not be formed and fed back. For most problems concerning the reciprocal interference of paths are due to local cycles. On the other hand, if the probability that an edge is taken is proportional to the increment of pheromone trail and inverse proportional to the edge cost, then a shortcut should be preferred from the exact time it is discovered by an ant. To cater on these problems with the ACO model, an alternative method for analysing the traffic flow and optimising the congestion in its path is introduced here. A water mass is unleashed at some high point. Gravity will make it to follow a path down until it cannot go down anymore. In Geology terms, when it rains in a mountain, water tries to find its own way down to the sea. Along the way, water erodes the ground and transforms the landscape, which eventually creates a riverbed. When a strong down slope is traversed by the water, it extracts soil from the ground in the way. This soil is deposited later when the slope is lower. Rivers affect the environment by reducing or increasing the altitude of the ground. If water is unleashed at all points of the landscape then the river form tends to optimize the task of collecting all the water and take it to the sea, which does not imply to take the shortest path from a given origin point to the sea. There are a lot of origin points to consider In fact, a kind of combined grouped shortest path is created in this case. The formation of tributaries and meanders is a consequence of this. However, if water flows from a single point and no other water source is considered, then the water path tends to provide the most efficient way to reduce the altitude (i.e., it tends to find the shortest path).

Instead of associating pheromone values to edges, the altitude values to nodes could be associated. Drops erode the ground (they reduce the altitude of nodes) or deposit the sediment (increase it) as they move. The probability of the drop to take a given edge instead of others is proportional to the gradient of the down slope in the edge, which in turn depends on the difference of altitudes between both nodes and the distance (i.e. the cost of the edge). At the beginning, a flat environment is provided, that is, all nodes have the same altitude. The exception is the destination node, which is a hole. Drops are unleashed at the origin node, which spread around the flat environment until some of them fall in the destination node. This erodes adjacent nodes, which creates new down slopes, and in this way the erosion process is propagated. New drops are inserted in the origin node to transform paths and reinforce the erosion of promising paths. After some steps, good paths from the origin to the destination are found. These paths are given in the form of sequences of decreasing edges from the origin to the destination. This method provides the following advantages with respect to ACO. On the one hand, local cycles are not created and reinforced because they would imply an ever decreasing cycle, which is contradictory. Though ants usually take into account their past path to avoid repeating nodes, they cannot avoid to be led by pheromone trails through some edges in such a way that nodes may be repeated in the next step. On the contrary, altitudes cannot lead drops to these situations. On the other hand, when a shorter path is found in RFD, the subsequent reinforcement of the path is fast, since the same origin and destination are concerned in both the old and the new path, the difference of altitude is the same but the distance is different. Hence, the edges of the shorter path necessarily have higher down slopes and are immediately preferred (in average) by subsequent drops. Moreover, the erosion process provides a method to avoid inefficient solutions: If a path leads to a node that is lower than any adjacent node (i.e., it is a blind alley) then the drop will deposit its sediment, which will increase the altitude of the node. Eventually, the altitude of this node will match the altitude of it neighbours, which will avoid that other drops fall in this node. Moreover, more drops could be cumulated in the node until the mass of water reaches adjacent nodes (a lake is formed). If water reaches this level, other drops will be allowed to cross this node from one adjacent node to another. Thus, paths will not be interrupted until the sediment fills the hole. This provides an implicit method to avoid inefficient behaviours of drops. As an improvement over the existing ACO model, the Traffic Congestion Forecasting Algorithms using the Pheromone Communication Model (PCM) could be enhanced to give more accurate and better results using the River Formation Dynamics (RFD). Further, a better optimisation of the congestion is expected as the RFD model accounts for the intrinsic characters missing in the ACO model.

## 4   Internet Of Things (IoT) and Intelligent Traffic System (ITS)

Smart City is the product of accelerated development of the new generation information technology and knowledge-based economy, based on the network combination of the Internet, telecommunications network, broadcast network, wireless broadband network and other sensors networks where Internet of Things technology (IoT) as its core. The main features of a smart city include a high degree of information technology integration and a comprehensive application of information resources. The Internet of Things is about installing sensors (RFID, IR, GPS, laser scanners, etc.) for everything, and connecting them to the internet through specific protocols for information exchange and communications, in order to achieve intelligent recognition, location, tracking, monitoring and management. With the technical support from IoT, smart city will have three features of being instrumented, interconnected and intelligent. The concept of Intelligent Transportaion Syetem (ITS) and IoT can be combined to give the best results on Traffic Congestion Optimisation. The idea is to deploy the sensors alongside the road units to count for the road side data. Then these observed data by the sensors would be sent to the local road agents installed across the road crossings joining a road unit. The road agent data will then be sent to the central server for the computation of the traffic congestion density of that particular road unit using the Traffic Congestion Optimisation Algorithm. Similarly, a simultaneous congestion index for all the road units would be calculated on the central server from the data received from the different road agents intsalled across different road units. The Global Positioning System (GPS) installed in the vehicles could then be routed to the path of minimum congestion density and the Traffic Light Cycle Timings (TLCTs) can synchroniously be handeled to meet the minimum traffic congestion. The ease of the inter-connectedness of all the 'Agents' of the traffic control system with IoT would increase the robustness in the traffic flow with the support of a self adaptive leaning algorithm for the congestion optimisation at the central server.

## 5 Conclusions

The present work gives a review on the various Evolutionary Algorithms (EAs) and Heuristic Models studied for the optimisation problems. Particularly, the Genetic Algorithms (GAs) involving the concepts of crossovers and mutations for optimising the objective function using various features, Swarm Particle Optimisation (SPO) using the bird flocking or fish schooling phenomenon to optimise their search paths, Ant Colony Optimisations (ACO) using the pheromone communication model, Bee Colony Optimisation (BCO) for the ride sharing- ride matching problems, Back Propagation Neural Networks (BPNN) using supervised learning were elaborated. Furthermore, the Random Walk Heuristic model for the state-space sequences with the parallelization concepts were discussed. Markov Chain Heuristic model for the finite states and known probabilities for state transition along with an introduction to the Large Step Markov Chains were pointed. As reviewed, EAs tends to optimise the problem using natural references, the Heuristic Models aims at more mathematical modifications of these naturally occurring optimisation phenomenon. The much unexplored River Formation Dynamics (RFD) was introduced in the regard of optimisation approach to the Traffic Congestion problem as an improvement over the ACO model. Later, the Internet of Things methodology for the Intelligent Traffic Control Systems (ITS) was introduced, which could be used for a robust, fast and smart traffic system.

## 6 Acknowledgements

We sincerely acknowledge the guidance by Dr. Gaurav Trivedi and Prof. S. Manjhi, Department of Electronics and Electrical Engineering, IIT Guwahati for the progress of the present research work. We thank our Ph.D. Mentor, Mr. Satyabrata Das, EEE Dept., IIT Guwahati who provided insight and expertise that greatly assisted the research of this paper.